\documentclass[twocolumn,aps,pre,floatfix,superscriptaddress,longbibliography]{revtex4-1}
\pdfoutput=1 %This is for the ArXiv submission only
\usepackage{amsmath,amssymb,eucal,graphicx,float,epstopdf}

\usepackage[colorlinks=true, urlcolor=blue, anchorcolor=blue, citecolor=blue,filecolor=blue,linkcolor=blue,menucolor=blue]{hyperref}

\newcommand{\sid}[1]{\textcolor{red}{\bf [#1]}}

\begin{document}
\title{Charged Aggregation}
\author{P. L. Krapivsky}
\affiliation{Department of Physics, Boston University, Boston, MA 02215, USA}
\affiliation{Santa Fe Institute, 1399 Hyde Park Road, Santa Fe, NM 87501, USA}
\author{S. Redner}
\affiliation{Santa Fe Institute, 1399 Hyde Park Road, Santa Fe, NM 87501, USA}

\begin{abstract}

  We introduce an aggregation process that begins with equal
  concentrations of positively and negatively `charged' monomers.
  Oppositely charged monomers merge to form neutral dimers.  These
  dimers are the seeds for subsequent aggregation events in which
  neutral clusters of necessarily even mass join irreversibly to form
  neutral aggregates of ever-increasing size.  In the mean-field
  approximation with mass independent reaction rates, we solve for the
  reaction kinetics and show that the concentration of clusters of
  mass $k$, $c_k(t)$, asymptotically scales as $A_k/t$, with $A_k$
  having a non-trivial dependence on $k$.  We also investigate the
  phenomenon of gelation in charged aggregation when the reaction rate
  equals the product of the two incident cluster masses.  Finally, we
  generalize our model to the case of three and more types of
  monomers.
  
\end{abstract}  
\maketitle

\section{Introduction}

Aggregation is a fundamental kinetic process in which clusters of
various masses irreversibly join to form clusters of ever-increasing
mass~\cite{Flory,friedlander2000smoke}.  If we denote a cluster of
mass $i$ by $C_i$, each reaction can be written as
\begin{align*}
C_i+C_j\mathop{\longrightarrow}^{K_{i,j}}C_{i+j}\,.
\end{align*}
Here $K_{i,j}$ is the reaction kernel, which specifies the rate at
which a cluster of mass $i$ (an $i$-mer) joins to a $j$-mer to form an
$(i+j)$-mer.  The basic observable in aggregation is the
time-dependent cluster-mass distribution whose nature depends on the
functional form of the reaction kernel.  In the approximation that all
reactants are perfectly mixed, the time dependence of the cluster mass
distribution is described by an infinite set of rate equations that
account for the change in the cluster concentrations due to reactions
with other clusters.

The reaction rates $K_{i,j}$ depend on the properties of the two
reacting clusters~\cite{Smo17,Chandra}.  For diffusion-controlled
reactions in three dimensions $K_{i,j}\sim (D_i+D_j)(R_i+R_j)$, where
$D_i$ and $R_i$ are the diffusion coefficient and the radius of a
cluster of mass $i$, respectively.  In turn, the cluster diffusion
coefficient is inversely proportional to its mass, which gives
$K_{i,j}\propto 2+(i/j)^{1/3}+(j/i)^{1/3}$~\cite{Chandra,OTB89,KRB}.
Because of the complicated form of this reaction rate, aggregation of
Brownian clusters is still unsolved~\cite{Colm12}.  However, a number
of idealized exactly soluble cases are known, including the constant
kernel ($K_{i,j}=$ const., which has the same homogeneity degree as
the Brownian kernel), the sum kernel ($K_{i,j}=i\!+\!j$), the product
kernel ($K_{i,j}=ij$), and a few other specialized
forms~\cite{Ziff,Leyvraz03}.  The constant-kernel case, in which the
reaction rates are independent of the cluster masses, is particularly
simple, and investigations of this toy model have helped to develop
the concept of scaling in aggregation~\cite{Ernst85a,Ernst88}.

\begin{figure}[ht]
\includegraphics[width=0.44\textwidth]{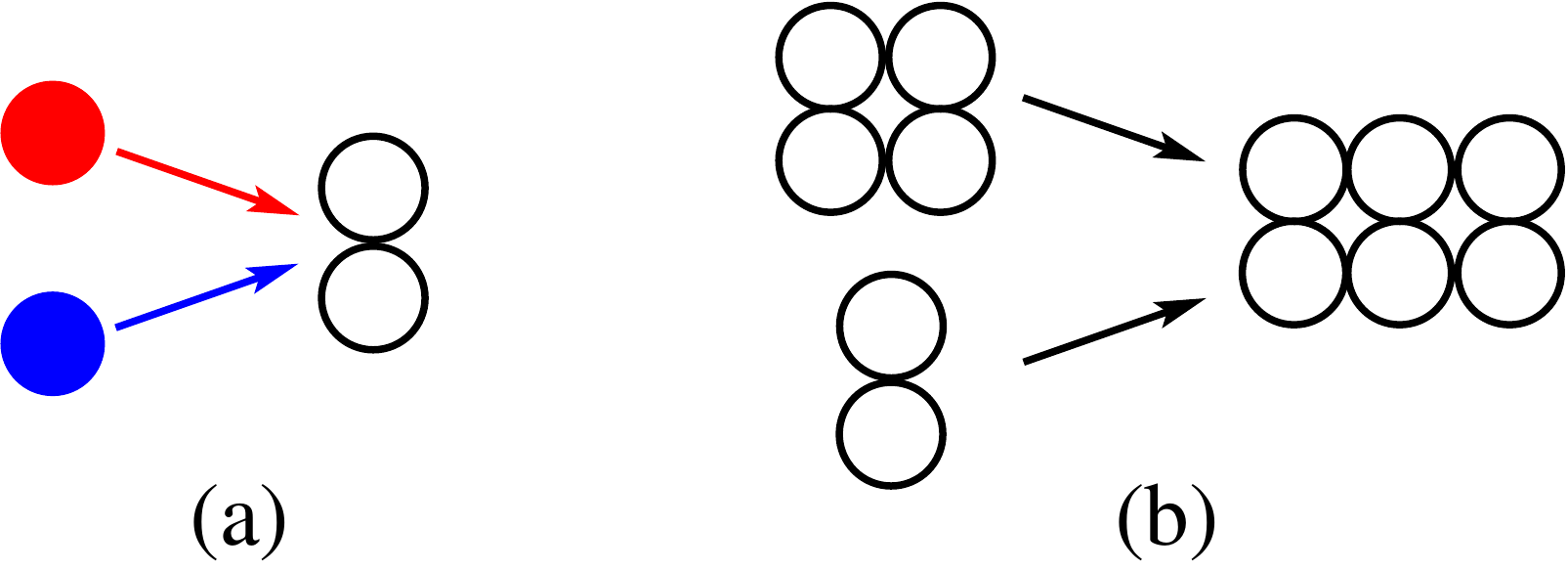}
\caption{ Charged aggregation: (a) Two oppositely charged monomers
  merge into a neutral dimer, (b) Two neutral clusters (here a dimer
  and a 4-mer) merge into a 6-mer.}
\label{fig:model}
\end{figure}

In this work, we investigate an extension of aggregation that begins
with equals concentrations of monomers of two types, $A$ and $B$, that
we label as positively charged and negatively charged.  There are no
physical electrostatic forces acting, but we invoke the label `charge'
to impose the constraint that only positively and negatively charged
monomers can merge to form dimers via the reaction
$[A]\oplus[B]\to [AB]$, while monomers of the same charge do not
interact.  Each dimer contains one $A$ and one $B$ monomer and thus
are neutral (Fig.~\ref{fig:model}(a)).

In addition to the interaction between oppositely charged monomers,
neutral clusters interact with a rate that is independent of their
masses.  Thus dimers constitute the seeds to generate neutral clusters
of ever-increasing masses.  Once a neutral dimer is created, it can
react only with other neutral clusters, and neutral clusters of mass
greater than or equal to 2 can interact among themselves to create
neutral clusters of the form $[(AB)^k]$ for all $k\geq 2$
(Fig.~\ref{fig:model}(b)). No other type of reactions occur.  For
electrically charged monomers, one should include repulsive
interactions between similar monomers and attractive interactions
between dissimilar monomers.  More importantly, electrostatic
interactions are long-ranged, and systems with long-range interactions
exhibit peculiar behaviors (see, e.e.,~\cite{Slava96, Ginzburg97} and
references therein).  In our modeling, we ignore all electrostatic
effects because our main interest is the role that the stoichiometry
of our model plays in the reaction kinetics.

One of the motivations for this work is to incorporate compositional
constraints on aggregation in a simple way.  In our charged
aggregation model, each aggregate necessarily consists of equal
numbers of $A$ and $B$ monomers.  However, one can envision that the
ratio of $A$s to $B$s in an aggregate is arbitrary, which defines a
richer class of models.  Physical realizations of aggregation that
involve more than one type of monomer have been observed
experimentally~\cite{Flory,von1980measurement,feder1984scaling}, and
models of aggregation with various compositional constraints have been
formulated~\cite{L83,Meakin_1986,LR86,LR87,meakin1988reaction}.  Our
modeling is focused on understanding the role of a particularly simple
compositional constraint on the aggregation kinetics.

One of our main results is that when the reaction rates are
independent of the cluster masses, $K_{i,j}=\text{const.}$, the
cluster mass distribution in charged aggregation is quite different
from that in classic aggregation with mass-independent reaction rates.
For charged aggregation, the density of monomers decays as $1/(1+t)$,
while the cluster-mass distribution has the asymptotic form
\begin{align}
c_k \sim t^{-1}\,e^{-\epsilon k}\,\frac{\Gamma\big(k-\frac{1}{2}\big)}{\Gamma(k+1)}
\end{align}
for any fixed $k$ when $t\gg 1$, where here $c_k$ denotes the
concentration of neutral clusters of mass $2k$ (containing $k$
positive and $k$ negative monomers).  As we shall discuss, the
exponent $\epsilon>0$ depends on the ratio of the monomer-monomer and
cluster-cluster merging rates.  In contrast, for classic aggregation,
the concentration of clusters of mass $k$ at time $t$, $c_k(t)$, is
\begin{align}
  \label{ck-classic}
  c_k(t) = \frac{1}{(1+t)^2}\left(\frac{t}{1+t}\right)^{k-1}
  \mathop{\longrightarrow}_{t\to\infty}~~ \frac{1}{t^2}\; e^{-k/t}
\end{align}
when all the $K_{i,j}$ are set equal to 2.  From this form, the
typical cluster mass grows linearly with time and the distribution of
cluster masses is effectively constant for masses smaller than the
typical mass.  The mass distributions of charged and classic
aggregation are quite different in the small-mass limit.

In Sec.~\ref{sec:MFT}, we investigate charged aggregation within the
mean-field framework in which the reactants are assumed to remain
perfectly mixed at all times.  We also assume that all reaction rates
are equal.  As we shall show, the primary difference between classic
aggregation and charged aggregation is that the latter is driven by a
time-dependent source of dimers. We also treat the more general
situation where the reaction rates between monomers is different than
the reaction rate between clusters.  In Sec.~\ref{sec:MFT-3}, we
generalize to an aggregation process where the elemental building
blocks are monomers of three types: $A$, $B$, and $C$.  The reaction
starts by the merging of three dissimilar monomers into `neutral'
trimers: $[A]\oplus[B] \oplus[C] \to [ABC]$.  Neutral clusters of mass
3 and greater then undergo binary aggregation.  We again employ the
mean-field framework and determine the cluster mass distribution when
the cluster merging rate is independent of the cluster masses.

\section{Two Monomer Types}
\label{sec:MFT}

\subsection{Equal Monomer and Cluster Reaction Rates}

We denote the density of monomers with positive charge as $m(t)$.  We
also assume that the density of positively and negatively charged
monomers are equal.  The time dependence of the monomer density
(either positively or negatively charged) is described by the rate
equation
\begin{equation} 
\label{m-eq}
\frac{dm}{dt}=-m^2\,,
\end{equation}
with solution, for the initial condition $m(0)=1$,
\begin{equation} 
\label{m-sol}
m = \frac{1}{1+t}\,.
\end{equation}

Let $c_k$ denote the concentration of neutral clusters of mass $2k$.
Under the assumption that neutral clusters react with constant and
mass-independent rates, the time dependence of the neutral cluster
densities is given by the rate equations
\begin{subequations}
\label{dck}
\begin{align} 
\label{dimers}
\frac{dc_1}{dt} &=-2c_1 c + m^2\,,\\[2mm]
\label{ck-eq}
\frac{dc_k}{dt}  &= \sum_{i+j=k}c_i c_j -  2c_k c\qquad k\geq 2\,,
\end{align}
\end{subequations}
where $c\equiv \sum_{k\geq 1} c_k$ is the total density of neutral
clusters.  A useful check of the correctness of these equations is to
compute the rate of change of the total mass density
\begin{align*}
  M\equiv m(t)+\sum_{k\geq 1} kc_k(t)\,.
\end{align*}
Adding Eq.~\eqref{m-eq} plus each of Eqs.~\eqref{dck} weighted by
their mass, it is immediate to see that $M$ is manifestly conserved.
Since we chose the initial monomer density to equal 1, the total mass
$M=1$.

To determine the individual cluster densities, it is necessary to
first solve for the total cluster density $c(t)$.  Summing
Eqs.~\eqref{dck}, we find that $c(t)$ satisfies the Riccati equation
\begin{equation}
\label{c-eq}
\frac{dc}{dt}=-c^2 + m^2=-c^2+\frac{1}{(1+t)^2}\,.
\end{equation}
This equation should be solved subject to initial condition $c(0)=0$.
While Riccati equations are generally unsolvable, some can be solved
by first guessing a particular solution, $c_*(t)$.  If such a solution
can be found, no matter how trivial, then the ansatz
$c(t)=c_*(t)+u(t)^{-1}$ reduces the Riccati equation to a linear
equation for $u(t)$ that can be solved by elementary
methods~\cite{Bender}.

\begin{figure}[ht]
\includegraphics[width=0.44\textwidth]{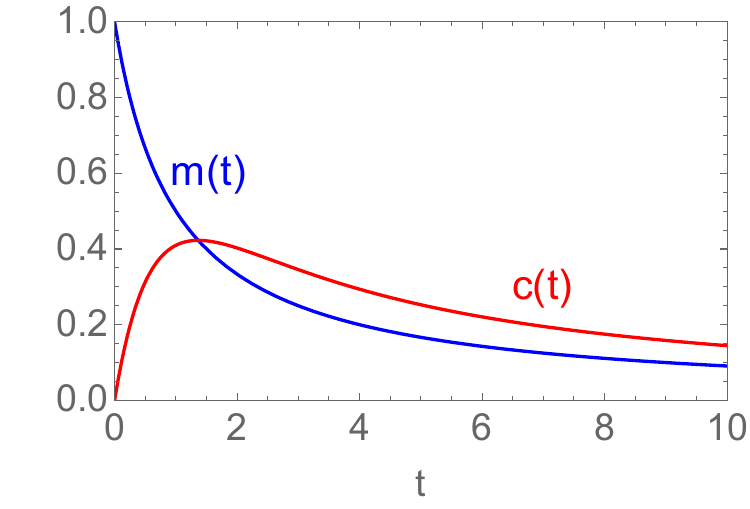}
\caption{Time dependence of the monomer density $m(t)$ from
  \eqref{m-sol} and the cluster density $c(t)$ from
  \eqref{c-sol}. Both densities decay as $t^{-1}$, and their ratio  $\frac{c(\infty)}{m(\infty)}$
  approaches  $\frac{1}{2}(\sqrt{5}+1)$. }
\label{fig:mc-t}
\end{figure}

The structure of the Riccati equation \eqref{c-eq} suggests seeking a
particular solution of the form $c_*=B/(1+t)$.  Substituting this
ansatz into Eq.~\eqref{c-eq}, it is straightforward to verify that
this ansatz indeed solves this equation if we choose
$B=(\sqrt{5}+1)/2$. Then the function $u(t)$ satisfies
\begin{equation*}
\frac{du}{dt} = 1 + \frac{\sqrt{5}+1}{1+t}\,u\,.
\end{equation*}
Solving this equation subject to the initial condition $c(0)=0$, the
full solution to \eqref{c-eq} is
\begin{equation}
\label{c-sol}
c(t) = \frac{2}{1+t}\,\frac{(1+t)^{\sqrt{5}}-1}{(\sqrt{5}-1)(1+t)^{\sqrt{5}} + \sqrt{5} + 1}\,.
\end{equation}
The densities $m(t)$ and $c(t)$ both asymptotically decay as $t^{-1}$
when $t\to\infty$ (Fig.~\ref{fig:mc-t}).  This behavior contrasts with
classic constant-kernel aggregation, where $m(t)$ asymptotically
decays as $t^{-2}$, while the cluster density asymptotically decays as
$t^{-1}$.  Intriguingly, for $t\to\infty$, the ratio of clusters to
monomers in charged aggregation approaches the golden ratio
$\frac{c(\infty)}{m(\infty)}=\frac{1}{2}(\sqrt{5}+1)$.

To determine the individual cluster densities $c_k(t)$, it is
expedient to introduce the generating function
\begin{equation*}
\mathcal{C}(z,t)\equiv \sum_{k\geq 1} c_k(t) z^k\,.
\end{equation*}
Multiplying each of Eqs.~\eqref{dck} by $z^k$ and summing over all
$k$, we recast the infinite system \eqref{dck} into the single
differential equation
\begin{equation}
\label{C-eq}
\frac{d \mathcal{C}}{dt}=\mathcal{C}^2 - 2c \mathcal{C} +\frac{z}{(1+t)^2}\,.
\end{equation}
As in the case of the generating function solution to classic
constant-kernel aggregation, it proves convenient to subtract
\eqref{C-eq} from \eqref{c-eq} to give the Riccati equation for
$y(z,t)=c(t)-\mathcal{C}(z,t)$:
\begin{equation}
\label{y-eq}
\frac{dy}{dt}=-y^2 +\frac{1-z}{(1+t)^2}\,,
\end{equation}
subject to the initial condition $y(z,0)=0$.  We solve this equation
by using the same approach that was used in solving Eq.~\eqref{c-eq}.
From this solution and also using the expression for $c(t)$ in
Eq.~\eqref{c-sol}, we finally obtain
\begin{align}
\label{GF-sol}
\mathcal{C}(z,t) &= \frac{2}{1+t}\,\frac{(1+t)^{\sqrt{5}}-1}{(\sqrt{5}-1)(1+t)^{\sqrt{5}} + \sqrt{5} + 1} \nonumber \\
&\qquad - \frac{2(1-z)}{1+t}\,\frac{(1+t)^{\zeta}-1}{(\zeta-1)(1+t)^{\zeta} + \zeta + 1}\,.
\end{align}
where $\zeta=\sqrt{5-4z}$.

Expanding \eqref{GF-sol} in powers of $z$, one can, in principle,
extract $c_k(t)$ for any $k\geq 1$.  However, the expression for $c_1$
is already cumbersome, and the expressions for $c_k$ for $k\geq 2$ are
even more so.  If we only want the asymptotic behavior, this may be
more easily obtained by substituting the ansatz $c_k=A_k/t$ into
\eqref{dck}; one may readily check that this substitution is self
consistent.  After straightforward steps, we find that the $A_k$
satisfy the recurrence
\begin{equation*}
%\label{A-rec}
\sqrt{5}\,A_k = \sum_{i+j=k}A_i A_j + \delta_{k,1}\,.
\end{equation*}
For $k=1$, we have $A_1=\frac{1}{\sqrt{5}}$. To obtain the general
solution for $A_k$, we introduce the generating function
$\mathcal{A}(z)\equiv \sum_{k\geq 1}A_k z^k$, multiply the above
recurrence by $z^k$, and sum over all $k$.  This gives the
quadratic equation for the generating function,
$\mathcal{A}^2-\sqrt{5}\mathcal{A}+z=0$, whose solution is
\begin{align*}
  \mathcal{A}(z)=\frac{\sqrt{5}}{2}\Big(1\pm \sqrt{1-\frac{4}{5} z}\Big)\,,
\end{align*}
where we must choose the negative sign before the square root to have
the correct behavior for $z\to 0$.  Expanding this expression in a
Taylor series gives
\begin{subequations}
\begin{align}
\label{ck-asymp}
c_k\simeq \frac{A_k}{t}\,,
\end{align}
with
\begin{align}
  A_k = \sqrt{\frac{5}{16\pi}}\;\frac{\Gamma\big(k-\frac{1}{2}\big)}{\Gamma(k+1)}\; \left(\frac{4}{5}\right)^{k}\,.
\end{align}
\end{subequations}
The mass distribution decays exponentially in $k$, with a
time-independent cutoff.  This result for $c_k$ is valid in the limit
$t\to\infty$ with $k$ fixed.

It is now instructive to compute the moments of the cluster-mass
distribution,
\begin{equation}
\label{Mn:def}
M_n \equiv \sum_{k\geq 1} k^n c_k\,.
\end{equation}
In this definition, we exclude the contribution of monomers for
convenience.  As we will show below, it is simpler to compute the
dependence of the moments on the monomer density rather than as a
function of time and then determine the asymptotic time dependence.
The exact expressions for the first three moments are
\begin{subequations}
\label{M123}
\begin{align} 
\label{M1-m}
M_1(m) &= 1-m\\
\label{M2-m}
M_2(m)  &=  \frac{2}{m}+1-3m + 4\ln m \\[2mm]
\label{M3-m}
M_3(m) &= \frac{(1-m)(6+24m+19m^2)}{m^2}\nonumber \\
        &\quad+ \frac{24}{m}\,\ln m +24\ln m +12(\ln m)^2
\end{align}
\end{subequations}

Equation \eqref{M1-m} is just the mass conservation statement.  To
obtain the second moment, we multiply each of
Eqs.~\eqref{dimers}--\eqref{ck-eq} by $k^2$ and sum these equations.
This gives the time dependence of the second moment:
\begin{subequations}
\begin{equation}
\label{M2-eq}
\frac{dM_2}{dt} = 2M_1^2+m^2\,.
\end{equation}
It is now helpful to use $\frac{dm}{dt}=-m^2$ and the mass
conservation statement $M_1=1-m$ to eliminate the time and express
$M_2$ as a function of $m$.  This gives
\begin{equation}
\label{M2-eq-m}
\frac{dM_2}{dm} = -2\big(1-m^{-1}\big)^2-1\,.
\end{equation}
\end{subequations}
The solution to this equation subject to the initial condition
$M_2(m\!=\!1)=0$ is just \eqref{M2-m}.

Similarly, the time dependence of the third moment is
\begin{subequations}
\begin{equation}
\label{M3-eq}
\frac{dM_3}{dt} = 6M_1 M_2+m^2\,.
\end{equation}
Once again, we eliminate the time in favor of $m$ to yield
\begin{equation}
\label{M3-eq-m}
\frac{dM_3}{dm} = -6(1-m)\left(\frac{2}{m}+1-3m + 4\ln m\right)-1\,.
\end{equation}
\end{subequations}
Solving this equation subject to the initial condition
$M_3(m\!=\!1)=0$ gives \eqref{M2-m}.

Since $m(t)\simeq t^{-1}$, the leading time dependence of the moments
in \eqref{M123} comes from the term with the most negative power of
$m$.  Thus we conclude that 
\begin{equation}
\label{M123-asymp}
M_1(t)\simeq 1, \quad M_2(t)\simeq 2t,  \quad M_3(t)\simeq 6t^2\,.
\end{equation}
Following the above line of reasoning and with some additional effort,
the time dependence of the general $n^{\rm th}$ moment as $t\to\infty$
is
\begin{equation}
\label{Mp-asymp}
M_n(t) \simeq n! \; t^{n-1}\,.
\end{equation}
This leading behavior coincides with the time dependence of the
moments in classic constant-kernel aggregation.  This equivalence
seems to stem from the fact that the reactions of neutral clusters in
charged aggregation is the same as reactions of all clusters in
constant-kernel aggregation.  The fact that neutral clusters are
created by a time-dependence source from the reaction of oppositely
charged monomers rather than being present in the initial state does
not seem to affect the long-time behavior of the moments.

There is a subtlety in the moments that deserves mention.  For a fixed
value of the mass $k$, the asymptotic behavior of the cluster density
is given by \eqref{ck-asymp}.  If \eqref{ck-asymp} remained valid for
all $k$, then all the moments $M_n(t)$ would decay as $t^{-1}$, since
$A_k$ decrease exponentially with mass and the sum
$ \sum_{k\geq 1} k^n A_k$ converges for all $n$.  Thus it is necessary
to take the limits $k\to\infty$ and $t\to\infty$ in the correct order.

\subsection{Distinct Monomer and Cluster Reaction Rates}

Because charged monomers are fundamentally distinct from neutral
clusters, it is natural to investigate the aggregation kinetics for
the situation in which the rate of cluster-cluster merging is set to
one, as before, but the monomer-monomer merging rate is set to
$\lambda$.  We now explore the consequences of this generalization
within the mean-field approximation.  The rate equations for the
monomer and  cluster densities now are (compare with
Eqs.~\eqref{m-eq} and \eqref{c-eq})
\begin{equation} 
\label{m-c1}
\frac{dm}{dt}=-\lambda m^2, \qquad \frac{dc}{dt} = - c^2+\lambda m^2\,.
\end{equation}
The rate equation for the the dimer density now is
\begin{align}
  \frac{dc_1}{dt} =-2c_1 c + \lambda m^2\,,
\end{align}
while the rate equations for the cluster densities with $k\geq 2$ are
again given by Eq.~\eqref{ck-eq}.

Solving the equations for $m(t)$, $c(t)$, and then $c_1(t)$, the
resulting expressions for the latter two quantities are extremely
cumbersome.  However, if we only want the asymptotic behavior, we may
perform the same analysis as given above to obtain the amplitude $A_k$
in the asymptotic expression for $c_k$, as well as the ratio of
clusters to monomers as a function of $\lambda$:
\begin{align} 
\label{Ak}
  \begin{split}
    &A_k = \left(\frac{4}{4+\lambda}\right)^{k}\frac{\sqrt{4+\lambda}}{4\sqrt{\pi \lambda}}\,\frac{\Gamma\big(k-\frac{1}{2}\big)}{\Gamma(k+1)}\,,\\[2mm]
&\frac{c(\infty)}{m(\infty)}=\frac{\sqrt{\lambda(4+\lambda)}+\lambda}{2}\,.
  \end{split}
\end{align}

The results for $c(t)$ and $c_1(t)$ simplify considerably for a number
of special cases.  For example, when $\lambda=\frac{1}{2}$, the above
ratio equals 1.  In this case, the cluster density becomes the
following rational function of time:
\begin{subequations}
\begin{equation}
\label{mc-2}
m = \frac{2}{2+t}\,, \quad c = \frac{2t}{2+t}\,\frac{12+6t+t^2}{24+12t+6t^2+t^3}\,,
\end{equation}
while the dimer density $c_1$ is
\begin{equation}
\label{c1-2}
c_1 = \frac{2}{3}\,\frac{A(t)+B(t)\ln(1+t/2)}{(2+t)(24+12t+6t^2+t^3)^2}\,,
\end{equation}
\end{subequations}
with
\begin{equation*}
\begin{split}
&A = t(480 + 384 t + 184 t^2 + 60 t^3 + 12 t^4 + t^5)\,, \\
&B=96 (2+t)^3\,.
\end{split}
\end{equation*}

\begin{figure}[ht]
\includegraphics[width=0.5\textwidth]{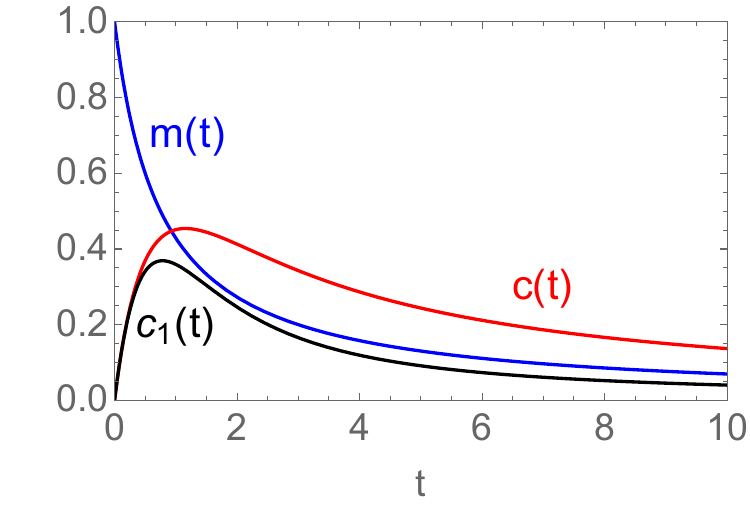}
\caption{The monomer density $m(t)$, the cluster density $c(t)$, and
  the dimers density $c_1(t)$ from Eqs.~\eqref{mcc-4/3} when
  $\lambda=\frac{4}{3}$.  All densities decay as
  $t^{-1}$. Asymptotically, the density ratios
  $\frac{c(\infty)}{m(\infty)}$ and $\frac{m(\infty)}{c_1(\infty)}$
  both equal 2.}
\label{fig:mcc}
\end{figure}

Another simple case is $\lambda=\frac{4}{3}$, where the ratio
$c(\infty)/m(\infty)$ now equals 2 (Fig.~\ref{fig:mcc}).  Here, the
cluster density is again a rational function of time
\begin{subequations}\label{mcc-4/3}
\begin{equation}
\label{mc-4/3}
m = \frac{3}{3+4t}\,, \quad c = \frac{12t}{3+4t}\,\frac{3+2t}{9+6t+4t^2}\,,
\end{equation}
and the dimer density $c_1$ is
\begin{equation}
\label{c1-4/3}
c_1 = \frac{3}{8}\,\frac{A(t)+B(t)\ln(1+4t/3)}{(3+4t)(9+6t+4t^2)^2}\,,
\end{equation}
\end{subequations}
with
\begin{equation*}
  A = t(540 + 504 t + 192 t^2 + 64 t^3), \quad B=27 (3+4t)^2\,.
\end{equation*}

\section{The Product Kernel}

We now investigate charged aggregation when the reaction kernel has
the product form $K_{i,j}=ij$, which leads to a finite-time gelation
transition.  At a critical gelation time an infinite cluster (gel
molecule) is born that gradually engulfs all the remaining finite-mass
clusters as $t\to\infty$

For generality, we assume that the reaction rate between monomers is
$\lambda$, so that its rate equation is the first of \eqref{m-c1},
with solution $m(t)=1/(1+\lambda t)$.  The density of cluster is now given by 
\begin{subequations}
  \label{dck-prod}
\begin{align} 
\label{dimers-prod}
\frac{dc_1}{dt} &=\lambda m^2 - c_1(1-m)\,,\\[2mm]
\label{ck-eq-prod}
\frac{dc_k}{dt}  &= \frac{1}{2}\sum_{i+j=k} i j c_i c_j -  k c_k (1-m)\qquad k\geq 2\,.
\end{align}
\end{subequations}
In the loss term in \eqref{ck-eq-prod}, we have used the fact that the
mass density of clusters, $\sum_{k\geq 1} kc_k$ equals $1-m$.

\begin{figure}[t]
\includegraphics[width=0.44\textwidth]{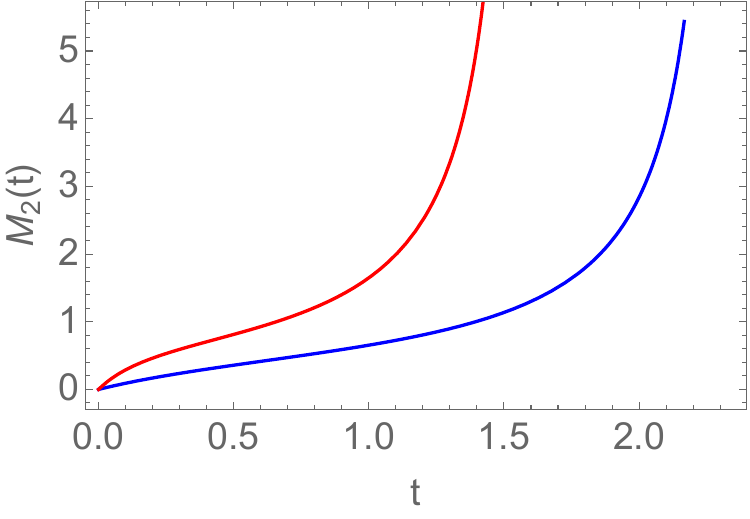}
\caption{The second moment $M_2(t)$ for product kernel charged
  aggregation for $\lambda=1$ (blue) and $\lambda=4$ (red).}
\label{fig:M2}
\end{figure}

To find the gelation transition, we study the time dependence of the
second moment $M_2(t)$.  From Eqs.~\eqref{dck-prod}, this second
moment satisfies
\begin{equation*}
%\label{M2-eq}
\frac{dM_2}{dt}=M_2^2 + \lambda m^2= M_2^2 +\frac{\lambda}{(1+\lambda t)^2}\,,
\end{equation*}
whose solution is
\begin{equation}
\label{M2-sol}
M_2 =  \frac{\Lambda}{1+\lambda t}\,\frac{1-(1+\lambda t)^{\sqrt{1-4/\lambda}}}{\Lambda-1-(1+\lambda t)^{\sqrt{1-4/\lambda}}}\,,
\end{equation}
where $\Lambda \equiv \frac{1}{2}(\lambda+\sqrt{\lambda(\lambda-4)})$.
This expression is manifestly real for $\lambda>4$ and it can be
recast into a real form for $0<\lambda\leq 4$.  For the specific cases
of $\lambda=1$ and $\lambda=4$, we find (Fig.~\ref{fig:M2})
\begin{equation}
\label{M2-sol-1}
M_2 =\begin{cases}
{\displaystyle  \frac{2}{1+t}\,\frac{1}{\sqrt{3}\,\cot\!\left[\frac{\sqrt{3}}{2}\ln(1+t)\right]-1}}&\lambda =1\,,\\[7mm]
{\displaystyle  \frac{2}{1+4t}\,\frac{\ln(1+4t)}{2-\ln(1+4t)}}& \lambda =4\,.
\end{cases}                                                      
\end{equation}

The second moment diverges at the gelation time, whose value is
obtained by setting the denominator in Eq.~\eqref{M2-sol} to zero.
This gives
\begin{equation}
\label{gel}
t_g = \frac{(\Lambda-1)^{1/\sqrt{1-4/\lambda}}-1}{\lambda}\,.
\end{equation}
For the special cases of $\lambda=1$ and $\lambda=4$, the gelation
time is $t_g=e^{2\pi/\sqrt{27}}-1=2.3508\ldots$ and
$t_g=\frac{1}{4} (e^2-1)=1.597264\ldots$, respectively.  The
limiting behaviors of the gel time are $t_g \to 1$ for
$\lambda\to \infty$ and $t_g\to \sqrt{\pi/4\lambda}$ for
$\lambda\to 0$.

From the rate equations \eqref{dimers-prod}--\eqref{ck-eq-prod}, the
time dependence of the total cluster density is
\begin{equation}
\label{c-eq-prod}
\frac{dc}{dt} = \lambda m^2 + \frac{g^2-(1-m)^2}{2}\,,
\end{equation}
where $g$ is mass in the gel phase.  This gel mass is defined via
\begin{equation}
M_1 = \sum_{k\geq 1} kc_k=1-m-g\,,
\end{equation}
where the sum is over finite-mass clusters.
When $t<t_g$, we integrate \eqref{c-eq-prod} with $g=0$ to yield
\begin{equation}
\label{c-sol-prod}
c = \frac{\ln(1+\lambda t)}{\lambda}- \frac{1-\lambda(1- t/2)}{1+\lambda t}\,t\,.
\end{equation}
This cluster density has a maximum for $t=\sqrt{2/\lambda}$.  In the
post-gel phase, $t>t_g$, we formally integrate \eqref{c-eq-prod} to
give
\begin{equation}
\label{c-post-prod}
c = \frac{\ln(1+\lambda t)}{\lambda}- \frac{1-\lambda(1- t/2)}{1+\lambda t}\,t+ \frac{1}{2}\int_{t_g}^t dt'\,g^2(t')\,.
\end{equation}

\section{Three types of monomers}
\label{sec:MFT-3}

\begin{figure}[ht]
\includegraphics[width=0.2\textwidth]{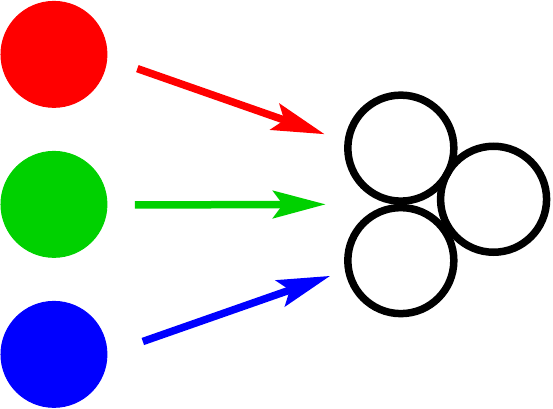}
\caption{Aggregation with 3 monomer species.  The elemental event
  where 3 distinct monomer types merge into a neutral trimer is
  shown.}
\label{fig:model3}
\end{figure}

Given the rich dynamics of charged aggregation with two types of
monomers, it is natural to generalize to the case of three types of
monomers, $A$, $B$, and $C$, that are each initially present with
equal densities.  In the same spirit as the two-monomer model, we
postulate that the only possible monomer reaction event is the merging
of three dissimilar monomers, $[A]\oplus[B] \oplus[C]\to [ABC]$, which
results in a neutral trimer (Fig.~\ref{fig:model3}).  Neutral clusters
of mass greater than or equal to 3 continue to merge to create neutral
clusters of the form $[(A B C)^k]$ with $k\geq 2$.  If we ascribe a
complementary color to each monomer species, the trimer is neutral
since it has no net color. 

Adopting the language from particle physics, we can think of monomers
as quarks with colors $A=$ red, $B=$ green, and $C=$ blue.  A baryon
is composed of three quarks, which must contain one monomer each of
red, green, and blue colors.  Hence, a trimer plays the role of an
elementary baryon.  We may also envision a more general stoichiometry
in which there are both quarks and antiquarks.  In the context of
particle physics, a quark-antiquark pair corresponds to a meson.  In
the framework of aggregation, one can imagine a rich range of
phenomena with both baryonic aggregation, mesonic aggregation, and
possibly mixed aggregation of baryons and mesons.

Returning to our minimal three-species model, the density $m$ of
monomers of each type decays according to
\begin{equation} 
\label{m-eq-3}
\frac{dm}{dt}=-m^3\,,
\end{equation}
whose solution, subject to $m(0)=1$, is
\begin{equation} 
\label{m-sol-3}
m = \frac{1}{\sqrt{1+2t}}\,.
\end{equation}
The density $c_1$ of trimers now satisfies
\begin{align} 
\label{trimers}
\frac{dc_1}{dt} =-2c_1 c + m^3\,,
\end{align}
while the density $c_k$ of clusters of mass $3k$ satisfies
Eq.~\eqref{ck-eq} for $k\geq 2$.

The time dependence of the total cluster density is accounted for by
the Riccati equation
\begin{equation}
\label{c-eq-3}
\frac{dc}{dt}=-c^2+(1+2t)^{-3/2}\,.
\end{equation}
While this equation is unsolvable, it is not difficult to determine
the relevant large-time behavior.  We first note that it is not
possible that all three terms in \eqref{c-eq-3} have the same time
dependence.  If one assumes that $c\sim t^{-\alpha}$, then the terms
in this equation are of order $t^{-(1+\alpha)}$, $t^{-2\alpha}$, and
$t^{-3/2}$, which can never be of the same order.  Thus we seek a
solution in which two of the three terms in \eqref{c-eq-3} are
dominant.  The only consistent solution arises when the terms on the
right-hand side are dominant, while the left-hand side is negligible.
With this assumption, we immediately find
\begin{equation}
\label{c-sol-3}
c \simeq  (1+2t)^{-3/4}\,.
\end{equation}
% Hereinafter we keep only the leading term and mention that the
% correction terms can be also computed.  For example, the next
% correction term to $c$ is given by
% \begin{equation*}
% c = (1+2t)^{-3/4}+\frac{3}{4}\, (1+2t)^{-1}+\ldots 
% \end{equation*}

By keeping the two dominant terms in \eqref{trimers}, we find that the
leading asymptotic behavior of the trimer density is simply $c_1=c/2$.
Using \eqref{ck-eq} we can then find the leading asymptotic behavior
of the densities $c_k$ for the first few $k$ values, from which we
deduce that all the $c_k$ are of the same order as $c$ itself.  Using
this fact, we therefore write
\begin{equation}
\label{ckA}
c_k = A_k\, c
\end{equation}
for any fixed $k$ and $t\to\infty$. Substituting this ansatz into
\eqref{ck-eq} and keeping only the leading terms gives the recurrence
\begin{equation*}
%\label{A-rec-3}
2A_k = \sum_{i+j=k}A_i A_j + \delta_{k,1}
\end{equation*}
whose solution is~\cite{KRB}
\begin{equation}
\label{ck-sol-3}
c_k = \frac{c}{\sqrt{4\pi}}\,\frac{\Gamma\big(k-\frac{1}{2}\big)}{\Gamma(k+1)}
\simeq \frac{c}{\sqrt{4\pi}}\,k^{-3/2} \qquad k\gg 1\,.
\end{equation}
% This distribution has an algebraic tail
% \begin{equation}
% \label{ck-asymp-3}
% c_k \simeq \frac{c}{\sqrt{4\pi}}\,k^{-3/2} \qquad (k\gg 1)
% \end{equation}

Let us now determine the asymptotic behavior of the moments $M_n$.
The first moment is given by Eq.~\eqref{M1-m} due to mass
conservation.  Following the same steps as those used for monomers
with two types of charges, the next two moments are
\begin{subequations}
\label{M23}
\begin{align} 
\label{M2-m-3}
M_2(m)  &=  (2-m^{-1})^2-m-2\ln m\,, \\[2mm]
\label{M3-m-3}
M_3(m) &= \frac{3}{2 m^4}-\frac{10}{m^3}+\frac{21}{m^2}- \frac{18}{m}+\frac{13}{2}-m \nonumber \\[1mm]
        &\qquad -6(1-m^{-1})^2\ln m\,.
\end{align}
\end{subequations}
We now substitute the asymptotic form $m(t)\simeq 1/\sqrt{2t}$ for the
monomer density into Eqs.~\eqref{M23} and find that the leading time
dependence of the moments are the same as in the case of two types of
monomers; that is, $M_n(t)\simeq n!\,t^{n-1}$.

If there are $n+1$ types of monomers with $n\geq 2$, the same
considerations as those used for the three-species model lead to
\begin{equation}
\label{mc-n}
m=(1+nt)^{-{1}/{n}}\,, \qquad c\simeq (1+nt)^{-{(n+1)}/{2n}}\,.
\end{equation}
The solution for the cluster-mass density is still given by
Eq.~\eqref{ck-sol-3}, but with $c$ now given by \eqref{mc-n}.  We also
find that the asymptotic time dependence of the moments is independent
of the number of monomer types.

\section{Summary and Discussion}

We investigated the kinetics in a model of `charged' aggregation, in
which the reaction begins with equal concentrations of positively
charged and negatively charged monomers.  Oppositely charged monomers
join to form neutral dimers, and neutral clusters of any mass greater
than or equal to 2 react freely with other neutral clusters to form
aggregates of ever-increasing size.  Within the mean-field
approximation, we obtained the time dependences of the concentration
of monomers and the concentration of clusters of any size.

At a qualitative level, charged aggregation is a version of classic
aggregation, but with a time-dependent source of dimers (effectively
the elemental constituents of charged aggregation) that is decaying
with time.  This mechanism leads to the densities of clusters of mass
$k$ decaying with time as $t^{-1}$ in the small-mass limit.  We also
explored the kinetics of charged aggregation with a product reaction
kernel.  We found that this model undergoes a conventional
second-order gelation transition at a gelation time than depends on
$\lambda$, the monomer-monomer reaction rate.

It should be worthwhile to explore the kinetics of charged aggregation
in a system of finite spatial dimension, where fluctuation effects
should play a significant role.  The simplest case and the one with
the largest departures from mean-field behavior is the case one
dimension.  There are two natural situations that may be worthwhile to
explore: (a) a spatially homogeneous system, (b) positive and negative
monomers initially occupying the positive and negative infinite
half-lines.  Another natural situation is a steady and spatially
localized monomer input in a $d$-dimensional space.

For the first scenario, charged aggregation involves a superposition
of a two-species reaction, the conversion of oppositely charge
monomers to neutral dimers, and single-species reactions, the merging
to neutral clusters of any size.  In one dimension, these two
constituent reactions have very different kinetics and their melding
could lead to unusual kinetics.

If the monomers of each type are spatially separated, then their
reaction is identical to the well-studied problem of two-species
annihilation with the same initially separated initial condition.  In
charged aggregation, the localized zone where monomers react leads to
a spatially localized and time-dependent source of dimers.  It is
natural to also treat a steady, but spatially localized source of
dimers in general spatial dimension. 

\bigskip
We thank Steen Rasmussen for a helpful conversation.  This work was
supported in part by the Santa Fe Institute.

\bibliography{references-CA}

\end{document}